\documentclass[conference]{IEEEtran}
\ifCLASSINFOpdf
\else
\fi
\usepackage{amsfonts}
\usepackage{amssymb}
\usepackage{amsmath}
\usepackage{graphicx}
\usepackage{subfigure}
\usepackage{algorithm}
\usepackage{algpseudocode}
\usepackage{cite}
\usepackage{multirow}
\usepackage{array}
\usepackage{multirow}
\usepackage{array}
\usepackage{soul}
\usepackage{slashbox}
\usepackage{algorithm}
\usepackage{algpseudocode}
\usepackage{booktabs}
\usepackage{threeparttable}
\usepackage{stfloats}

\begin{document}

\title{An SMDP-based Resource Management Scheme for Distributed Cloud Systems }
\author{\authorblockN{Jiadi~Chen, Hang~Long, Qiang~Zheng, Minyao Xing, Wenbo~Wang}
\authorblockA{
Wireless Signal Processing and Network Lab, Key laboratory of Universal Wireless Communication, \\ Ministry of Education,
Beijing University of Posts and Telecommunications, Beijing, China \\
Email: chenjd@bupt.edu.cn
 }} \maketitle

\begin{abstract}

In this paper, the resource management problem in geographically distributed cloud systems is considered. The Follow Me Cloud concept which enables service migration across federated data centers (DCs) is adopted. Therefore, there are two types of service requests to the DC, i.e., new requests (NRs) initiated in the local service area and migration requests (MRs) generated when mobile users move across service areas. A novel resource management scheme is proposed to help the resource manager decide whether to accept the service requests (NRs or MRs) or not and determine how much resources should be allocated to each service (if accepted). The optimization objective is to maximize the average system reward and keep the rejection probability of service requests under a certain threshold.  Numerical results indicate that the proposed scheme can significantly improve the overall system utility as well as the user experience compared with other resource management schemes.

\end{abstract}

\footnotetext[1]{ This work is supported in part by the Fundamental Research Funds for the Central Universities (No.2014ZD03-02), National Key Technology R\&D Program of China (2014ZX03003011-004) and China Natural Science Funding (61331009).  }

\section{Introduction}

The booming of bandwidth-intensive mobile applications and ever growing mobile user number have brought great challenges for today's mobile cloud systems~\cite{MCC}. One of the performance bottlenecks lies in the fact that most current mobile cloud systems are highly centralized. The fast growing of mobile cloud computing business is calling for geographically distributed cloud infrastructures, i.e., federated data centers (DCs)~\cite{DC1}, to relieve the heavy load of the central server and improve user experience.

Except the specialized cloud providers like Google, the evolution of mobile network architecture has promoted the decentralization of cloud systems as well. Toward the fifth generation (5G)~\cite{5G} of wireless broadband, emerging paradigms such as network function virtualization (NFV)~\cite{nfv} can help to realize a  flat and intelligent mobile network embraced with cloud technology. For example, in a mobile system adopted the concept of C-RAN~\cite{C-RAN}, scattered cloud resource blocks within a certain geographical area can be aggregate into a virtual cloud resource pool. Federated resource pools in multiple geographical areas can thus form a distributed cloud system.

The Follow Me Cloud (FMC) concept, which enables service migration across federated DCs following the mobility of mobile terminals (MTs), is widely accepted in distributed cloud systems~\cite{FMC}. Enjoying service from the geographically nearest DC can always get a pleasant user experience since a local DC can minimize the end-to-end delay between MTs and cloud servers.

In this paper, a geographically distributed cloud system which adopt the FMC concept is considered. Multiple DCs take charge of respective service areas and each DC is equipped with a resource manager (RM).  A MT can initiate a new request (NR) to the local DC in its resident service area. When the MT wanders to another service area during service period, a service migration request (MR) will be send to the destination DC. The corresponding RM then makes a trade off between the user perceived quality and incurred system cost to decide whether the service should be migrated. RMs also make decisions on how much resources should be allocated to each accepted service request (NR or MR).

The resource management problem described previously is formulated as an constrained semi-Markov Decision Process (SMDP). Our work is inspired by~\cite{FMC1} and~\cite{FMC2}. Authors in~\cite{FMC1} introduce an analytical model for FMC concept and in~\cite{FMC2}, the service migration procedure is modeled using MDP. Other relevant works on service migration in cloud systems include~\cite{fmc3},~\cite{fmc4}, etc. In these previous works, the service model is described from the perspective of MTs, and the resource allocation problem is not considered. In our work, the decision making approach is described from the perspective of the overall system, and the objective is to improve the overall system utility.
To solve the SMDP, the value iteration algorithm and Q-learning algorithm are employed to obtain the optimal policy. Compared with other schemes, the SMDP-based resource management scheme can significantly increase the average system reward and reduce the rejection probability of service requests.

The remainder of this paper is organized as follows. The system model and service migration procedure is described in Section \ref{model}. In Section \ref{problem}, the resource management problem is formulated as an SMDP. The solution to the problem and the numerical results are presented in Section \ref{solution}. Finally, the conclusion is drawn in Section \ref{conclusion}.

\section{System Model} \label{model}

\begin{figure}
\centering
\subfigure[Migration request accepted.]{ \label{fmc1} 
\includegraphics[width=0.4\textwidth]{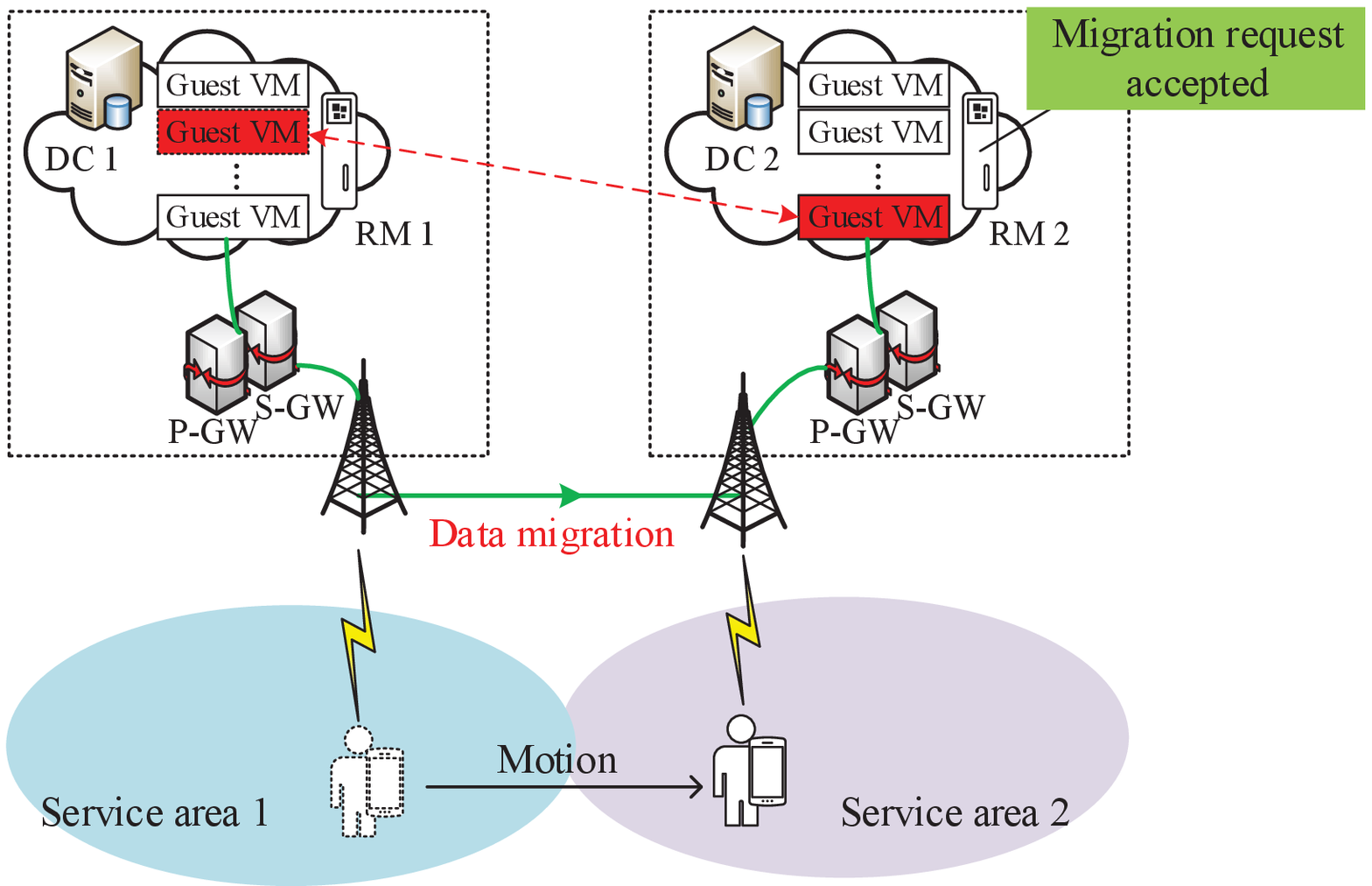}}
\subfigure[Migration request rejected.]{ \label{fmc2} 
\includegraphics[width=0.4\textwidth]{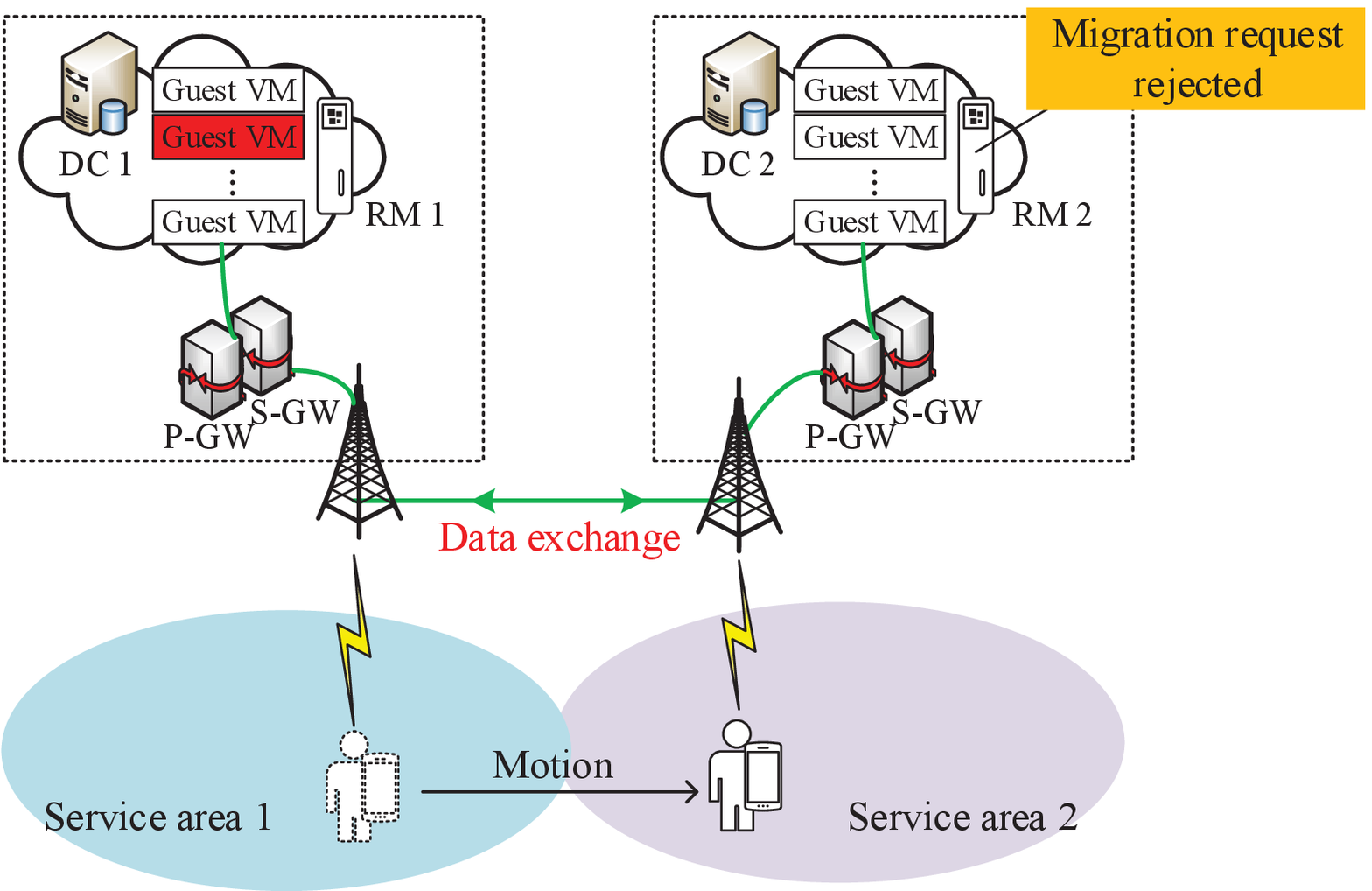}}
\vspace{-5pt}
\caption{Service migration diagram. } \label{2user} 
\vspace{-20pt}
\end{figure}

We consider a typical 3GPP cellular network covered by heterogeneous wireless access nodes, e.g., macro base station (BS), femtocell node~\cite{femto}, WLAN access point, etc. Each hexagonal cell is equivalent to a service area and be assigned a DC.  For simplicity, we assume that each cell is covered by a single macro BS which is collocated with a DC. When the MT enters the coverage area of a BS, it enters the service area of the attached DC equivalently. Each DC has a resource pool which contains $B$ units of resources. MTs are offered resources (computation or storage) represented in the form of virtual machines (VMs). After a service request is accepted, a guest VM is constructed in the DC and the RM will allocate resources to the VM for running the service.

When a MT moves from one service area to another during service period,  a service MR will be sent to the destination DC. If the MR is accepted, as shown in Fig. \ref{fmc1}, the corresponding VM in the original DC will be released and a new VM will be constructed in the destination DC. During migration process, service data in the original VM needs to be transmitted to the new VM via backhaul. Another case is that the MR is rejected, as Fig. \ref{fmc2} shows, the VM in the original DC will be maintained and the MT will receive service via an additional wired link between the serving DC and the current connected base station.

The arrival process of NRs is modeled as a Poisson process with the rate of ${\lambda _n}$. Let $p_m$ denote the cross-area movement rate of MTs and the service time is assumed to be geometrically distributed with mean ${1 \mathord{\left/{\vphantom {1 {\left( {1 - \mu } \right)}}} \right.\kern-\nulldelimiterspace} {\left( {1 - \mu } \right)}}$. Therefore, the arrival process of MRs is also poissonian with the rate of
\vspace{-5pt}
\begin{equation}
{\lambda _m} = {\lambda _n}\left( {1 - \mu } \right){p_m}.
\vspace{-10pt}
\end{equation}

\begin{figure}
  \includegraphics[width=0.48\textwidth]{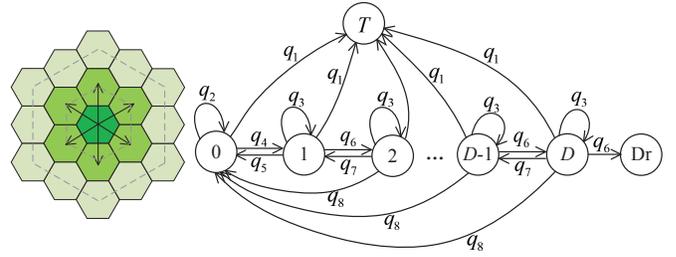}\\
  \vspace{-20pt}
  \caption{Transition diagram of service distance, \footnotesize {$q_1=\mu$, ${q_2} = \left( {1 - \mu } \right)\left( {1 - {p_m}\overline {p_r^m}} \right)$,
${q_3} = \left( {1 - \mu } \right)\left( {1 - {p_m} + 2\overline {p_r^m} /6} \right)$, ${q_4} = \left( {1 - \mu }\right){p_m}\overline {p_r^m} $, ${q_5} = \left( {1 - \mu } \right){p_m}\left( {1 - 5\overline {p_r^m} /6} \right)$, ${q_6} = \left( {1 - \mu } \right)3{p_m}\overline {p_r^m} /6$, ${q_7} = \left( {1 - \mu } \right){p_m}\overline {p_r^m} /6$, ${q_8} = \left( {1 - \mu } \right){p_m}\left( {1 - \overline {p_r^m} } \right)$.}}\label{move}
\vspace{-20pt}
\end{figure}

A six-directional random walk mobility model~\cite{mobility_model} is used to characterize the user movement. When a MT moves between two adjacent service areas, the previous distance and current distance between the MT and its serving DC (defined by hop count among service areas) are denoted as $d_p$ and $d_c$, respectively. The transition probability of service distance can be obtained by
\begin{equation}
\small{
\Pr [{d_c}|{d_p}] = \left\{ {\begin{array}{*{20}{l}}
{\left( {1 - \mu } \right)\left( {1 - {p_m}\overline {p_r^m} } \right),}&{{d_p} = {d_c} = 0}\\
{\left( {1 - \mu } \right){p_m}\overline {p_r^m} ,}&{{d_p} = 0,{d_c} = 1}\\
{\left( {1 - \mu } \right)\left( {1 - \dfrac{{5\overline {p_r^m} }}{6}} \right),}&{{d_p} = 1,{d_c} = 0}\\
{\left( {1 - \mu } \right)\dfrac{{3{p_m}\overline {p_r^m} }}{6},}&{\begin{array}{*{20}{l}}
{{d_p} > 0,}\\
{{d_c} = {d_p} + 1}
\end{array}}\\
{\left( {1 - \mu } \right)\dfrac{{{p_m}\overline {p_r^m} }}{6},}&{\begin{array}{*{20}{l}}
{{d_p} > 0,}\\
{{d_c} = {d_p} - 1}
\end{array}}\\
{\left( {1 - \mu } \right)\left( {1 - {p_m} + \dfrac{{2{p_m}\overline {p_r^m} }}{6}} \right),}&{{d_c} = {d_p} > 0}\\
{\left( {1 - \mu } \right){p_m}\left( {1 - \overline {p_r^m} } \right)}&{{d_p} > 0,{d_c} = 0}
\end{array}} \right.,}
\end{equation}
where ${\overline {p_r^m} }$ denotes the average rejection probability of MRs. $d_c=0$ indicates that the MT is receiving service form the local DC. The transition diagram of service distance is illustrated in Fig. \ref{move}, where $T$ denotes the state that the service has been finished. $D$ is the maximum allowable service distance. When service distance exceeds $D$, the service will be interrupted and this situation is represented by state $\rm{Dr}$. Based on the transition diagram, we can obtain the probability distribution of each service distance value, i.e., ${\mathbb{P}_d} = \left( {\Pr \left[ {{d_c} = 1} \right],\Pr \left[ {{d_c} = 2} \right], \cdots ,\Pr \left[ {{d_c} = D} \right]} \right) \in {\left[ {0,1} \right]^D}$.

\section{Problem Formulation} \label{problem}

In this section, the service model described previously is formulated as an constrained SMDP.
From the perspective of a service area, the system state $s$ describes the resource occupation in the DC and an event occurs in the system, i.e.,
\begin{equation}
s \in \left\{ {{s^L},{s^R},e} \right\} \in \mathcal{S},\quad e \in \mathbf{E} = \left\{ {\mathbf{A,T,M}} \right\},
\end{equation}
where $\mathcal{S}$ is the system state space. ${s^L}$ and ${s^R}$ are defined as
\begin{equation}
{s^L} = \left\{ {s_1^L,s_2^L, \cdots ,s_C^L} \right\},\quad {s^R} = \left\{ {s_1^R,s_2^R, \cdots ,s_C^R} \right\} .
\end{equation}
$s_c^L$ denotes the number of local MTs (MTs that are currently in this service area) which occupies $c$ unit resources of this DC. Similarly, $s_c^R$ denotes the number of remote MTs (MTs which are located in other service areas but still receive service from this DC) that occupies $c$ unit resources. Total amount of occupied resources of this DC is $\sum\nolimits_{c = 1}^C {c\left( {s_c^L + s_c^R} \right)}  \le B$, $C$ denotes the maximum amount of resources that can be allocated to a single service. $e$ represents an event occurs in the system and the event set $\mathbf{E}$ is described as follows:
\begin{itemize}
  \item $\mathbf{A} = \left\{ {{A^n},{A^m}} \right\}$. ${A^n}$ and ${A^m}$ denote the arrival of a NR and a MR to this DC, respectively.
  \item $\mathbf{T} = \left\{ {T_c^L,T_c^R|c \in \left\{ {1, \cdots ,C} \right\}} \right\}$. $T_c^L$ denotes the finish and departure of a local service (service to a local MT) which occupies $c$ unit resources. Similarly, $T_c^R$ denotes the finish of a remote service (service to a remote MT) which occupies $c$ unit resources. 
  \item $\mathbf{M} = \left\{ {M_c^L,M_c^R|c \in \left\{ {1, \cdots ,C} \right\}} \right\}$. $M_c^L$ denotes the cross-area movement of a local MT that occupies $c$ unit resources. Similarly, $M_c^R$ denotes the cross-area movement of a remote MT that occupies $c$ unit resources. 
\end{itemize}

The occurrence time points of a sequence of events are called decision epochs which are indexed by $k \in \left\{ {1,2, \cdots } \right\}$ in chronological order.  At each decision epoch, the RM chooses an action $a$ from the action space $\mathcal{A}_s$ which is defined as
\begin{equation}
{\mathcal{A}_s} = \left\{ {\begin{array}{*{20}{l}}
{\left\{ {0,1, \cdots ,C} \right\},}&{e \in \left\{ {{A^n},{A^m}} \right\}}\\
{ \{- 1\},}&{{\rm{otherwise}}}
\end{array}} \right..
\end{equation}
$a=0$ indicates that the request is rejected by RM. $a=c$ indicates that the request is accepted and $c$ units of resources are allocated to this service. In other cases,  the RM need not to make decisions but update the resource consumption (denoted by $a=-1$) in the system state. 

Based on the system state $s$ and the corresponding action $a$, the system reward function can be evaluated as
\vspace{-5pt}
\begin{equation}
r\left( {s,a} \right) = g\left( {s,a} \right) - \int_0^{y\left( {s,a} \right)} {d\left( {s,a} \right)dt} ,
\vspace{-5pt}
\end{equation}
$g\left( {s,a} \right)$ is the lump sum income the system gains immediately after action $a$ is taken. $d\left( {s,a} \right)$ is the cost rate function which indicates the per unit time cost during service period and $y\left( {s,a} \right)$ is the expected sojourn time of system state until next decision epoch. The lump sum income function $g(s,a)$ can be expressed as
\vspace{-10pt}
\begin{equation}
\small{
g\left( {s,a} \right) = \left\{ {\begin{array}{*{20}{l}}
{{G_t}}&{e \in {\mathbf{T}}}\\
{{G_m}\left( {1 - \overline {p_r^m} } \right)}&{e \in {\mathbf{M}}}\\
{ - {C_l},}&{e = {A^n},a = 0}\\
{ - {C_m},}&{e = {A^m},a \ne 0}\\
{ - \dfrac{{{C_d}}}{2}\Pr \left[ {{d_c} = D} \right],}&{e = {A^m},a = 0}\\
{0,}&{{\rm{otherwise}}}
\end{array}} \right..}
\vspace{-5pt}
\end{equation}
$G_t$ and $G_m$ denote the system income for finishing a service and accomplishing a service migration, respectively. ${{C_l}}$ is the system loss caused by rejecting an NR and $C_m$ is the overhead incurred by service data migration from previous serving DC to the migration destination DC. Finally, $C_d$ is the system loss due to service interruption.

We take the end-to-end delay between MT's connected BS and its serving DC as the main QoS factor related to the service migration. This part of delay can influence the service response time and is proportional to the service distance. Another considered factor is the system resource occupation cost. Therefore, the $d(s,a)$ can be expanded as
\vspace{-5pt}
\begin{equation}
d\left( {s,a} \right) = \sum\limits_{c = 1}^C {\left( {{\omega _d}s_c^R\overline d + {\omega _o}{C_r}c\left( {s_c^L + s_c^R} \right)} \right)}.
\vspace{-5pt}
\end{equation}
${C_r}$ denotes the one unit resource occupation price per unit time. $\omega_d$ and $ \omega_o$ ($\omega_d+\omega_o=1$) are weighting factors indicating the relative importance between service delay and resource occupation cost. ${\overline d}$ denotes the average service distance which can be calculated by $\overline d = \sum\nolimits_{d = 1}^D {d\Pr \left[ {{d_c} = d} \right]}$.

The state duration function $y(s,a)$ which denotes the expect time length between two consecutive decision epochs, is the reciprocal of event rate $\gamma \left( {s,a} \right)$, i.e.,
\vspace{-5pt}
\begin{equation}
\gamma \left( {s,a} \right) = y{\left( {s,a} \right)^{ - 1}} = {\lambda ^n} + {\lambda ^m} + \sum\limits_{c = 1}^C {\left( {\mu  + {p_m}} \right)\left( {s_c^L + s_c^R} \right)}.
\vspace{-5pt}
\end{equation}

The RM chooses actions according to a certain policy which is defined as $\Omega  = \left( {{\delta _1}\left( s \right),{\delta _2}\left( s \right), \cdots } \right)$. ${\delta _k}\left( s \right) = a$ is the action decision rule at the $k$-th decision epoch. In this paper, we consider stationary policies only, which remain constant at different decision epochs, i.e., $\Omega  = \left( {\delta \left( s \right),\delta \left( s \right), \cdots } \right)$. Given a feasible unichain policy $\Omega $, the induced state transition process can form a Markov chain with transition probabilities of
\vspace{-5pt}
\begin{equation}
\small{
{p^\Omega }\left( {s'|s,\delta \left( s \right)} \right) = \left\{ {\begin{array}{*{20}{l}}
{\dfrac{{{\lambda ^n}}}{{\gamma \left( {s',\delta \left( s \right)} \right)}},}&{e' = {A^n}}\\
{\dfrac{{{\lambda ^m}}}{{\gamma \left( {s',\delta \left( s \right)} \right)}},}&{e' = {A^m}}\\
{\dfrac{{s_c^L\mu }}{{\gamma \left( {s',\delta \left( s \right)} \right)}},}&{e' = T_c^L}\\
{\dfrac{{s_c^R\mu }}{{\gamma \left( {s',\delta \left( s \right)} \right)}},}&{e' = T_c^R}\\
{\dfrac{{s_c^L{p_m}}}{{\gamma \left( {s',\delta \left( s \right)} \right)}},}&{e' = M_c^L}\\
{\dfrac{{s_c^R{p_m}}}{{\gamma \left( {s',\delta \left( s \right)} \right)}},}&{e' = M_c^L}
\end{array}} \right.,}
\vspace{-5pt}
\end{equation}
where $s,s' \in \mathcal{S}$ and $e'$ is the event element in state $s'$.

The average reward optimality criterion is used in this model. Therefore, the optimization objective of the SMDP is to achieve the optimal policy satisfying
\begin{equation}
\begin{array}{*{20}{l}}
{\mathop {\max }\limits_{\Omega  \in P} }&{\bar r = \mathop {\lim }\limits_{K \to \infty } {\mathbb{E}_\Omega }\left[ {\dfrac{1}{K}\sum\limits_{k = 0}^K {\dfrac{{{r_k}\left( {s,a} \right)}}{{{y_k}\left( {s,a} \right)}}} } \right]}\\
{{\rm{s}}.{\rm{t}}.}&{\overline {{p_r}}  = \mathop {\lim }\limits_{K \to \infty } {\mathbb{E}_\Omega }\left[ {\dfrac{1}{K}\sum\limits_{k = 0}^K {\left( {{\omega ^n}{{\overline {p_r^n} }^k} + {\omega ^m}{{\overline {p_r^m} }^k}} \right)} } \right] \le \rho }
\end{array}.
\end{equation}
$\mathcal{P}$ is the set containing all feasible policies. $\rm{\mathbb{E}}_\Omega[*]$ represents the expectation value of quantity * under policy $\Omega$. ${{r_k}\left( {s,a} \right)}$ and ${{y_k}\left( {s,a} \right)}$ are the reward function and state duration function in the $k$-th time period. ${{{\overline {p_r^n} }^k}}$ and ${{{\overline {p_r^m} }^k}}$ are average rejection probabilities of NRs and MRs at the $k$-th decision epoch, respectively. $\omega_n$ and $\omega_m$ are relevant importance factors and $\omega_n +\omega_m=1$. $\rho  = {\omega ^n}\hat p_r^n + {\omega ^m}\hat p_r^m$ is the threshold of rejection probability of service requests, where ${\hat p_r^n}$ and ${\hat p_r^n}$ are maximum allowable rejection probabilities of NRs and MRs, respectively.  By introducing the Lagrange multiplier $\beta$, the above constrained optimization problem can be converted to an unconstrained one as
\begin{equation}\label{objective}
\begin{array}{*{20}{l}}
{\mathop {\max }\limits_{\Omega  \in P} \quad \overline {{r_\beta }}  = \mathop {\lim }\limits_{K \to \infty } {\mathbb{E}_\Omega }\left[ {\dfrac{1}{K}\sum\limits_{k = 0}^K {\dfrac{{{r_k}\left( {s,a} \right)}}{{{y_k}\left( {s,a} \right)}}} } \right]}\\
{\quad \quad \quad \quad \quad \quad  - \beta \mathop {\lim }\limits_{K \to \infty } {\mathbb{E}_\Omega }\left[ {\dfrac{1}{K}\sum\limits_{k = 0}^K {\left( {{\omega ^n}{{\overline {p_r^n} }^k} + {\omega ^m}{{\overline {p_r^m} }^k}} \right)} } \right].}
\end{array}
\end{equation}
The optimal policy which satisfies the above equations can be obtained by solving the following Bellman equations~\cite{puterman} recursively, i.e.,
\vspace{-5pt}
\begin{equation}\label{bellman}\small{
\begin{array}{*{20}{l}}
{V\left( s \right) = \mathop {\max }\limits_{a \in {\cal A}} \left\{ {\mathop {{\rm{ }}{r_\beta }}\limits_{}^{} \left( {s,a} \right) - \theta y\left( {s,a} \right)} \right.}\\
{\quad \quad \quad \quad \quad \quad \left. { + \sum\limits_{s' \in {\cal S}} {{p^\Omega }\left( {s'|s,a} \right)V\left( {s'} \right)} } \right\},\forall s \in {\cal S}.}
\end{array}}
\vspace{-5pt}
\end{equation}
$\theta$ and $V(s)$ are called the average system gain and potential function of state $s$. $r_\beta(s,a)$ is the Lagrange reward function which is given by
\vspace{-5pt}
\begin{equation}\label{rbeta}
{r_\beta }\left( {s,a} \right) = g\left( {s,a} \right) - \int_0^{y\left( {s,a} \right)} {d\left( {s,a} \right)} dt - \beta f\left( {s,a} \right),
\vspace{-5pt}
\end{equation}
$f(s,a)$ is the constrain function which is given by
\begin{equation}
\small{
f\left( {s,a} \right) = \left\{ {\begin{array}{*{20}{l}}
{{\omega ^n},}&{e = {A^n},a = 0}\\
{{\omega ^m},}&{e = {A^m},a = 0}\\
{0,}&{{\rm{otherwise}}}
\end{array}} \right..}
\end{equation}

If there exists a policy $\Omega^*$ satisfying (\ref{bellman}), it is called the optimal policy and we have $\theta^* = \overline{r_\beta}^*$.  $\theta^* $ is the maximum average system gain in (\ref{bellman}) corresponds to the optimal policy $\Omega^*$. $\overline{r_\beta}^*$ is the maximum average gain which satisfies (\ref{objective}). 

\section{Problem Solution and Performance Evaluation} \label{solution}

\subsection{Solution to the Constrained SMDP}

In this paper, the VIA is used for obtaining the optimal policy, before which the SMDP has to be transformed to an equivalent discrete-time model as follows.
\begin{equation}\label{r}
{\tilde r_\beta }\left( {s,a} \right) \equiv {{{r_\beta }\left( {s,a} \right)} \mathord{\left/
 {\vphantom {{{r_\beta }\left( {s,a} \right)} {y\left( {s,a} \right)}}} \right.
 \kern-\nulldelimiterspace} {y\left( {s,a} \right)}},\forall s \in {\cal S},
\end{equation}
\begin{equation}
{\tilde p^\Omega }\left( {s'|s,a} \right) \equiv \left\{ {\begin{array}{*{20}{l}}
{{{\eta {p^\Omega }\left( {s'|s,a} \right)} \mathord{\left/
 {\vphantom {{\eta {p^\Omega }\left( {s'|s,a} \right)} {y\left( {s,a} \right),}}} \right.
 \kern-\nulldelimiterspace} {y\left( {s,a} \right),}}}&{s \ne s'}\\
{{{1 + \eta \left[ {{p^\Omega }\left( {s'|s,a} \right) - 1} \right]} \mathord{\left/
 {\vphantom {{1 + \eta \left[ {{p^\Omega }\left( {s'|s,a} \right) - 1} \right]} {y\left( {s,a} \right)}}} \right.
 \kern-\nulldelimiterspace} {y\left( {s,a} \right)}},}&{s = s'}
\end{array}} \right.,
\end{equation}
All quantities with ``$ \sim $" denote the corresponding ones in the transformed model. Then we can employ the VIA~\cite{puterman} directly with a given value of $\beta$.

\begin{algorithm}

\caption{ \quad The Q-learning VIA} \label{qlvia}
\begin{algorithmic}

\State 1. \small{Set $\beta = \beta^1$, $\beta^1$ is an arbitrary number greater than 0. $\quad$Specify $\varepsilon  > 0$ and set $n=1$.}
\State 2. \small{Substitute $\beta^{n}$ into (\ref{r}) as
\begin{equation}\label{1}
{\tilde r_{{\beta ^n}}}\left( {s,a} \right) \equiv {{{r_{{\beta ^n}}}\left( {s,a} \right)} \mathord{\left/
 {\vphantom {{{r_{{\beta ^n}}}\left( {s,a} \right)} {y\left( {s,a} \right)}}} \right.
 \kern-\nulldelimiterspace} {y\left( {s,a} \right)}},\forall s \in {\cal S}
\end{equation}
Solve for policy $\Omega^{n}$ in (\ref{bellman}) with a reward function given by (\ref{1}) via VIA.}

\State 3. \small{Calculate the system steady state distribution ${\pi ^{{\Omega ^n}}}$ under policy $\Omega^{n}$. Then calculate
\begin{equation}
{\overline {p_r}}^{\Omega _{{\beta ^n}}^ * } = \sum\limits_{s \in \mathcal{S}} {p_\infty ^{\Omega _{{\beta ^n}}^*}\left( s \right)f\left( {s,{\delta _{{\beta ^n}}}\left( s \right)} \right)} .
\end{equation}}

\State 4. \small{Let ${\Delta ^n} = \rho - {\overline {p_r}}^{\Omega _{{\beta ^n}}^ * }$, if ${\Delta ^n} < \varepsilon$ when $n \ge 2$, go to step 5. Otherwise update $b^n$ by
\begin{equation}
    {{\beta}^{n + 1}} = {{\beta}^n} + \frac{\alpha}{n}{\Delta ^n},
\end{equation}
where $\alpha$ is the step size which can be revised during the iteration process. Then go to step 2.}

\State 5. \small{Take ${\Omega ^n}$ as the optimal policy and stop.}
\end{algorithmic}
\end{algorithm}

The system state transition matrix under policy $\Omega$ is denoted as ${\mathbb{P}^\Omega } = \left( {{p^\Omega }\left( {s'|s,a} \right)|s,s' \in \mathcal{S}} \right) \in {\mathbb{R}^{M \times M}}$, where $M$ is the size of the system state space.  For a feasible $\rho$, there exists a $\beta^*$ satisfying
\begin{equation}\label{=}
{\overline {p_r}}^{\Omega _{{\beta ^*}}^*} = \sum\limits_{s \in {\cal S}} {p_\infty ^{\Omega _{{\beta ^*}}^*}\left( s \right)} f\left( {s,{\delta _{{\beta ^*}}}\left( s \right)} \right) = \rho,
\vspace{-5pt}
\end{equation}
$\Omega _{{\beta ^*}}^* = {\delta _{{\beta ^ * }}}\left( s \right)$ denotes the policy obtained by solving (\ref{bellman}) with $\beta^*$. Assume that $\beta^-$ is smaller than $\beta^*$ and $\beta^+$ is larger than $\beta^*$, then we have
\vspace{-10pt}
\begin{equation}\label{>}
{\overline {p_r}}^{\Omega _{{\beta ^ - }}^*} = \sum\limits_{s \in {\cal S}} {p_\infty ^{\Omega _{{\beta ^ - }}^*}\left( s \right)} f\left( {s,{\delta _{{\beta ^ - }}}\left( s \right)} \right) > \rho ,
\end{equation}
\vspace{-10pt}
\begin{equation}\label{<}
{\overline {p_r}}^{\Omega _{{\beta ^ + }}^*} = \sum\limits_{s \in {\cal S}} {p_\infty ^{\Omega _{{\beta ^ + }}^*}\left( s \right)} f\left( {s,{\delta _{{\beta ^ + }}}\left( s \right)} \right) < \rho.
\vspace{-5pt}
\end{equation}
$\Omega _{{\beta ^-}}^* = {\delta _{{\beta ^ - }}}\left( s \right)$ and $\Omega _{{\beta ^+}}^* = {\delta _{{\beta ^ + }}}\left( s \right)$ are policies obtained by solving (\ref{bellman}) with $\beta^-$ and $\beta^+$, respectively.

When solve (\ref{bellman}), we set $\beta$ as a arbitrary positive value and employ the VIA to get a temporary optimal policy $\Omega_{\beta}^*$. Then the expected rejection probability ${\overline {p_r}}^{\Omega _{{\beta}}^*}$ can be calculated with (\ref{=}). If ${\overline {p_r}}^{\Omega _\beta ^*} \ne \rho$, then we adjust the value of $\beta$ according to (\ref{>}) and (\ref{<}). Thus repeatedly, the value of $\beta$ can converges to $\beta^*$ with arbitrarily small error. This approach is referred to as the Q-learning VIA and the detailed flow is shown in \textbf{Algorithm \ref{qlvia}}.

\begin{figure}
  \centering
  \includegraphics[width=0.4\textwidth]{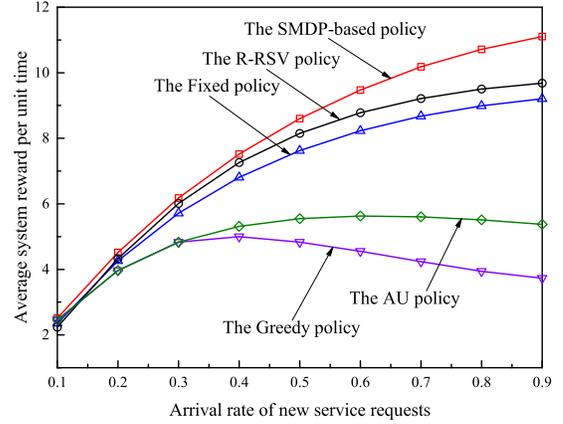}\\
  \vspace{-10pt}
  \caption{Average per unit time reward of the system under different policies.}\label{reward}
\end{figure}
\begin{figure}
 \vspace{-10pt}
\centering
  \includegraphics[width=0.4\textwidth]{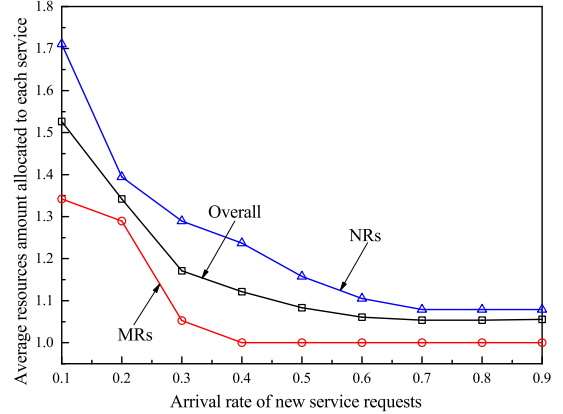}\\
  \vspace{-10pt}
  \caption{Average resource amount allocated to each service under the optimal policy.}\label{rn}
  \vspace{-15pt}
\end{figure}

\subsection{Numerical Results and Analysis}

In this subsection, the performance of the SMDP-based resource management scheme is evaluated. The optimal policy $\Omega^*$ obtained by employing the Q-learning VIA is compared with four reference baselines. Baseline 1 refers to a short-sighted policy obtained by employing the greedy algorithm~\cite{greedy} (called Greedy policy). Baseline 2 and 3 are two simple and straight allocation methods, one of which allocates all available resources in the DC to each service request (called AU policy) and another allocates a fixed amount of resources for all service requests (called Fixed policy). The last baseline refers to the resource reservation scheme proposed in~\cite{reserve} (referred to as the R-RSV policy) which reserves a small portion of the resources only for MRs. 

\begin{figure}
\centering
\subfigure[Average rejection probability of NRs.]{ \label{nr} 
\vspace{-20pt}
\includegraphics[width=0.4\textwidth]{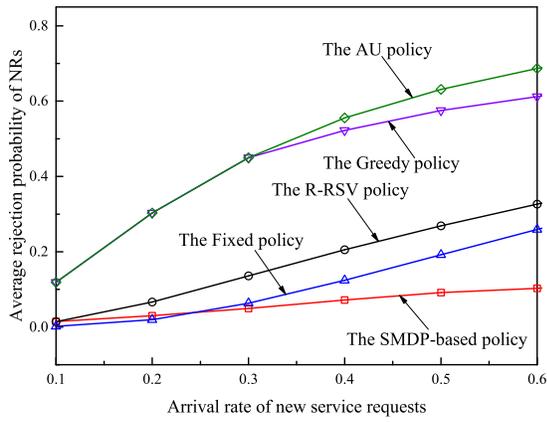}} 
\vspace{-5pt}
\subfigure[Average rejection probability of MRs.]{ \label{mr} 
\includegraphics[width=0.4\textwidth]{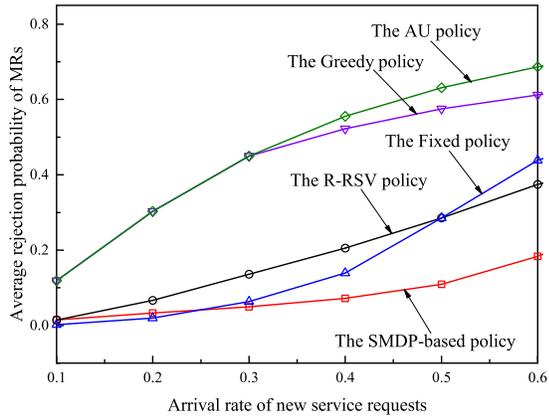}}
\caption{Average rejection probability of service requests under different policies.} \label{drop} 
\vspace{-20pt}
\end{figure}

The average per unit time reward of the system under different policies is presented first. As Fig. \ref{reward} shows, the SMDP-based policy outperforms other baselines on average system reward, especially when the service requests are intensively arriving. The R-RSV policy achieves a fine performance as well for it reserves some resources for migrated services, thus the rejection probability of MRs can be reduced.
Fig. \ref{rn} illustrates the average resource amount the RM allocates to each service request when the SMDP-based resource management scheme is adopted. We can see that the average resource amount allocated to an NR is higher than the average resource amount allocated to a MR. With the increase of NR arrival rate, the average resource amount allocated to each service request decreases to ensure that the DC can serve more MTs.

The rejection probabilities of NRs and MRs under different policies are illustrated in Fig. \ref{drop}. It can be seen that the SMDP-based policy can significantly reduce the rejection probabilities of both NRs and MRs. By adjusting the value of threshold $\rho$, the rejection probability of service requests can be controlled within a certain scope.
Therefore, we can conclude that for distributed cloud systems which support service migration, the proposed resource management scheme can significantly improve the overall system utility as well as the user perceived quality.

\vspace{-5pt}
\section{Conclusion} \label{conclusion}
\vspace{0pt}

In this paper, the resource management problem in geographically distributed cloud systems which adopt the FMC concept is considered. An SMDP-based admission control and resource allocation scheme is proposed to help RMs make decisions on whether to accept the service requests (NRs or MRs) or not and determine the amount of resources allocated to each accepted service. The optimization objective is to maximize the average system reward and keep the rejection probability of service requests under a given threshold. To determine the value of the Lagrange multiplier, the Q-learning algorithm is used and then the VIA is employed to obtain the optimal policy.  Numerical results indicate that the proposed resource management scheme can improve the system reward and reduce the rejection probability of service requests meanwhile.

\end{document}